\documentclass[10pt,journal,twocolumn]{IEEEtran}

\ifCLASSINFOpdf
  \usepackage[pdftex]{graphicx}
\else
  \usepackage[dvips]{graphicx}
\fi
\usepackage{algorithm}
\usepackage{algorithmic}
\usepackage{support-caption} 
\usepackage{subcaption}
\usepackage{silence}
\usepackage{comment} 
\usepackage{amsmath}
\usepackage{multirow}
\usepackage[T1]{fontenc}
\usepackage{tabularx}
\usepackage{tabu}
\usepackage{listings}
\usepackage{multirow}
\usepackage{xcolor}
\usepackage{listings}
\usepackage{caption}
\usepackage{tikz}
\usepackage{lipsum}
\usepackage{kantlipsum}
\usepackage{wasysym}
\usepackage{amssymb}
\usepackage{pifont}
\newcommand{\xmark}{\ding{55}}
\newcommand{\cmark}{\ding{51}}%

\newif\ifcommentson
\commentsontrue

\usepackage[pscoord]{eso-pic}

\begin{document}
\bstctlcite{IEEEexample:BSTcontrol}

%
\title{SRPerf: a Performance Evaluation Framework\\for IPv6 Segment Routing}

\author{Ahmed Abdelsalam, %
Pier Luigi Ventre, %
Carmine Scarpitta, %
Stefano Salsano\\%
Andrea Mayer, %
Pablo Camarillo, %
Francois Clad, %
Clarence Filsfils %
%

\vspace{2ex}
\textbf{\\Version 3 - March 2020 - Submitted paper under review}
\vspace{-4ex}
}


%

%


\maketitle

\begin{abstract}
Segment Routing is a form of loose source routing. It provides the ability to include a list of instructions (called \emph{segments}), in the packet headers. The Segment Routing architecture has been first implemented with the MPLS dataplane and then, quite recently, with the IPv6 dataplane (SRv6). IPv6 Segment Routing (SRv6) is a promising solution to support advanced services such as Traffic Engineering, Service Function Chaining, Virtual Private Networks, and Load Balancing. The SRv6 data-plane is supported in many different software forwarding engines including the Linux kernel and VPP software router, as well as in hardware devices. In this paper, we present SRPerf, a performance evaluation framework for software and hardware implementations of SRv6. SRPerf is able to perform different benchmarking tests such as throughput and latency. For throughput tests, we use the \textit{Partial Drop Rate} (PDR) to characterize a system under test
. The architecture of SRPerf can be easily extended to support new benchmarking methodologies as well as different SRv6 implementations. We have used SRPerf to evaluate the performance of the SRv6 implementation in the Linux kernel and in VPP. SRPerf is a valuable tool in the context of software forwarding engines where new features can be added at fast pace, as it helps experimenters to validate their work. In particular, we describe how we have leveraged SRPerf to validate the implementation of some SRv6 behaviors that were missing or wrongly implemented in the Linux kernel mainline. 
\end{abstract}

%
\IEEEpeerreviewmaketitle

\vspace{0.5ex}


\vspace{-2ex}
\section{Introduction}
\label{sec:introduction}


\IEEEPARstart{S}egment Routing is a network architecture based on the loose Source Routing paradigm (\cite{filsfils2015segment, rfc8402}). The basic concepts proposed in \cite{filsfils2015segment} have been elaborated and refined in the RFC 8402 \cite{rfc8402} which has recently completed its standardization process in the IETF (July 2018). In the SR architecture, a node can include an ordered list of instructions in the packet headers. These instructions steer the forwarding and the processing of the packet along its path in the network. The single instructions are called \textit{segments} and each segment can enforce a topological requirement (e.g. pass through a node or an interface) or a service requirement (e.g. execute an operation on the packet). Each segment is encoded by a Segment IDentifier (SID).


The SR architecture is supported by two different data-plane implementations: MPLS (SR-MPLS) and IPv6 (SRv6), in which SIDs are respectively encoded as MPLS labels and IPv6 addresses. Several use cases and requirements for Segment Routing have been collected in a number of documents. In \cite{rfc7855}, the main use cases identified are: IGP based tunneling (i.e. to support VPN services), Fast ReRoute (FRR), Traffic Engineering (further classified in a number of more specific use cases). A set of Resiliency use cases is described in \cite{rfc8355}. The Segment Routing use cases for IPv6 networks are considered in \cite{rfc8354} with a set of exemplary deployment environments for SRv6: Small Office, Access Network, Data Center, Content Delivery Networks and Core Networks. 

SR-MPLS has been the first instantiation of the SR architecture to be rolled out, which allowed to partially leverage the SR benefits (\cite{davoli2015} and \cite{cianfrani2016}), while the recent interest and developments are focusing on SRv6. The SRv6 implementations have drawn a lot of attention to researchers from academia and industry, as witnessed by the publication of several research activities 
\cite{ventre2019survey}. 
A strong open source ecosystem is supporting SRv6 advances. In particular, we want to mention \cite{srnet}, \cite{srorg} and the ROSE project \cite{rose}. Moreover, the data-plane implementations of SRv6 have been supported in many different routers implementations including: open-source software routers such as the Linux kernel and the Vector Packet Processing (VPP) platform~\cite{fd-io-vpp}, and hardware implementations from different network vendors~\cite{ietf-6man-segment-routing-header}. Finally, large scale deployments of SRv6 in production networks have been recently announced \cite{draft-matsushima-spring-srv6-deployment-status}. The list of announced SRv6 deployments expands several network segments such as service providers and data center networking. 
In this work, the SRv6 data-plane is our main focus and we aim at enable the benchmarking of the different SRv6 data-plane implementations. 

The introduction of SRv6 in ISP networks requires the assessment of its non-functional properties like scalability and reliability. Hence, the availability of a realistic performance evaluation framework for SRv6 is of fundamental importance. A measurement platform 
should allow scaling up to the current transmission line rates. Ideally, it should be available for re-use on any commodity hardware. To the best of our knowledge, there are no such open source performance measurements tools nor works that provide a complete performance evaluation for SRv6. There are some works that reports partial experiments for the SRv6 implementations, like~\cite{lebrun2017reaping,lebrun2017implementing,ahmedperformance} and the report published at~\cite{csit-report}. \cite{lebrun2017reaping} and~\cite{lebrun2017implementing} are early evaluations reporting the performance of the very first implementations of SRv6. \cite{ahmedperformance} provides an early implementation of a performance framework for Linux and reports the performance of some SRv6 behaviors and pointed out few performance issues without providing any solution to these issues. \cite{csit-report} focuses on VPP forwarding in general and reports the performance of few SRv6 behaviors. Considering the interest in performance analysis of SRv6 and the fact that these works do not provide a complete analysis of the available platforms and supported SRv6 behaviors, we advocate the need of an open source reference platform and complete analysis of currently implemented SRv6 behaviors. 

The design of a performance evaluation tool for data-plane implementations of forwarders is a very challenging task~\cite{intel-cisco-report} as they are required to forward packets at an extremely high rate using a limited CPU budget to process each of these packets. The IETF has defined the guidelines and the tests for benchmarking forwarders implementations~\cite{rfc2544}. The tests include: throughput, latency, jitter and frame loss rate. Throughput is the most commonly used measure for forwarders implementations~\cite{rfc1242}. It is defined as the maximum rate at which all received packets are forwarded by the device and often reported in number of packet per second (pps). RFC defines different variations of the throughput including No-Drop Rate (NDR), Maximum Receive Rate (MRR) and Partial Drop Rate (PDR)~\cite{csit-report}.

In this paper we present SRPerf, a performance evaluation framework for software and hardware implementations of SRv6. SRPerf is a modular framework supporting the performance evaluation of packet forwarding in the Linux kernel and in VPP. It can also be extended to support new forwarders. In addition to SRv6, SRPerf supports the performance evaluation of plain IP forwarding. It reports different throughput measures such as NDR, PDR and MRR. The current design relies on TRex~\cite{trex-cisco} as a traffic generator. The framework can be easily extended to support other packet generators. SRPerf is open-source and publicly available at~\cite{srperf}. 
The main contributions of this work are the following:

\begin{itemize}
	\item Realization of performance evaluation framework for software and hardware implementations of SRv6; current implementation supports Linux kernel and VPP as forwarders;
	\item Implementation of a generic PDR finder algorithm which allows to estimate with an user defined precision the PDR of a system under test;
	\item Performance evaluation of SRv6 forwarding behaviors supported both by the Linux kernel and by VPP;
	\item Improved performance of existing cross-connect behaviors (namely \textit{End.DX6} and \textit{End.X}) in the Linux kernel;
	\item Implementation of \textit{End.DT4} in the Linux kernel;
\end{itemize}
 
The paper is structured as follows: Section~\ref{sec:srv6-support} presents the SRv6 support in the Linux kernel and in VPP. The design of SRPerf and the evaluation methodology are described in Section~\ref{sec:framework}. 
Section~\ref{sec:performance} explains the testbed and presents the experiments we have performed. We also elaborate on two use cases, showing how we have leveraged SRPerf to benchmark the implementation of a new forwarding behavior and to identify and solve performance issues of existing SRv6 implementations. We report on the related works in Section~\ref{sec:related}. We draw some conclusions and highlight directions for future work in Section \ref{sec:conclusions}.

\section{SRv6 software implementations}
\label{sec:srv6-support}

In this section, we firstly provide an overview of the SRv6 networking programming concepts and then we review the status of the software open source implementations, namely Linux kernel (Sec. \ref{sec:srv6-linux}) and VPP router (Sec. \ref{sec:srv6-vpp}). A deeper introduction to SRv6 technology can be found in \cite{ventre2019survey}. Table~\ref{tab:support} shows the support of the SRv6 networking programming concepts and its extensions \cite{ietf-spring-srv6-network-programming,filsfils-spring-srv6-net-pgm-insertion,ietf-spring-sr-service-programming,id-srv6-mobile-uplane} in the Linux kernel and in VPP software router.

\begin{table}[htbp]
\caption{SRv6 support in Linux kernel and in VPP.}
\vspace{-1ex}
\label{tab:support}
\centering
\begin{tabular}{|c|l|c|c|c|}
\hline
\multirow{2}{*}{Category} & \multirow{2}{*}{Behavior}& \multirow{2}{*}{Linux}& \multirow{2}{*}{VPP} &\multirow{2}{*}{Measured} \\
& &  & & \\
\hline
\multirow{6}{*}{Headend}& H.Insert          & \cmark & \cmark & \cmark\\
\cline{2-5}
&H.Insert.Red      &  &  & \\
\cline{2-5}
&H.Encaps          & \cmark & \cmark & \cmark\\
\cline{2-5}
&H.Encaps.Red      &  &\cmark    &\\
\cline{2-5}
&H.Encaps.L2       & \cmark & \cmark  & \cmark\\
\cline{2-5}
&H.Encaps.L2.Red   &  &\cmark   &\\
\hline
\multirow{2}{*}{Endpoint}&End               & \cmark & \cmark & \cmark\\
\cline{2-5}
&End.T             & \cmark & \cmark  & \cmark\\
\cline{2-5}
(no-decap)&End.X             & \cmark & \cmark & \cmark\\
\hline
&End.DT4           & \cmark & \cmark  & \cmark\\
\cline{2-5}
&End.DT6           & \cmark & \cmark & \cmark\\
\cline{2-5}
&End.DT46          &  &   &\\
\cline{2-5}
Endpoint&End.DX2           & \cmark & \cmark & \cmark\\
\cline{2-5}
\multirow{2}{*}{(decap)}&End.DX4           & \cmark & \cmark  & \cmark\\
\cline{2-5}
&End.DX6           & \cmark & \cmark & \cmark\\
\cline{2-5}
&End.DX2V          &  &  &\\
\cline{2-5}
&End.DT2U          &  &   &\\
\cline{2-5}
&End.DT2M          &  &   &\\
\hline
\multirow{5}{*}{Binding SID}&End.B6.Insert     & \cmark & \cmark  &\\
\cline{2-5}
&End.B6.Insert.Red &  &  &\\
\cline{2-5}
&End.B6.Encaps     & \cmark & \cmark &\\
\cline{2-5}
&End.B6.Encaps.Red &  &\cmark    &\\
\cline{2-5}
&End.BM            &  &  &\\
\hline
\multirow{3}{*}{Proxy}&End.AS            &  & \cmark &\\
\cline{2-5}
&End.AD            &  & \cmark &\\
\cline{2-5}
&End.AM            &  & \cmark  &\\
\hline
\multirow{6}{*}{Mobile user-plane}& T.M.Tmap          &  & \cmark &\\
\cline{2-5}
&End.M.GTP4.E      &  & \cmark &\\
\cline{2-5}
&End.M.GTP4.D      &  & \cmark  &\\
\cline{2-5}
&End.GTP6.D.Di      &  & \cmark  &\\
\cline{2-5}
&End.M.GTP6.E      &  & \cmark &\\
\cline{2-5}
&End.M.GTP6.D      &  & \cmark &\\
\hline
\end{tabular}
\end{table}

\vspace{-1ex}

SRv6 has drawn a lot of interest since its introduction at IETF. This interest of the stakeholders as well as the trend of SDN has lead to wide range of SRv6 support both in software forwarders, such as Linux kernel and VPP, and hardware routers~\cite{ietf-6man-segment-routing-header}. These implementations have been revised several times to keep up with the evolution of the SRv6 network programming concepts \cite{ietf-spring-srv6-network-programming,filsfils-spring-srv6-net-pgm-insertion,ietf-spring-sr-service-programming} and \cite{id-srv6-mobile-uplane}.

SRv6 defines a new type of IPv6 routing extension header known as Segment Routing Header (SRH)~\cite{ietf-6man-segment-routing-header}. The SRH contains an ordered list of segments, which implements an SR policy. Each segment identifier (SID) is a 128-bit that has the form of an IPv6 address. A dedicated field, referred to as \textit{Segments Left}, is used to maintain the pointer to the active SID of the Segment List. 

The \textit{SRv6 Network Programming} model \cite{ietf-spring-srv6-network-programming} defines two different sets of SRv6 behaviors, known as \textit{SR policy headend} and \textit{endpoint} behaviors. \textit{SR policy headend} behaviors steer received packets into the SRv6 policy matching the packet attributes. Each SRv6 policy has a list of SIDs to be attached to the matched packets. On the other hand, an SRv6 \textit{endpoint} behavior, also known as behavior associated with a SID, represents a function to be executed on packets at a specific location in the network. Such function can be a simple routing instruction, but also any advanced network function (e.g., firewall, NAT). \textit{SR policy headend} behaviors are executed in the SR source node (also known as Headend node), while \textit{endpoint} behaviors in SR Segment Endpoint nodes. \textit{Endpoint} behaviors are further classified as \textit{decap} and \textit{no-decap} whether or not they perform decapsulation of the SRH. Transit nodes can be SR-capable or SR-incapable, as they do not need to inspect the SRH since the destination address of the packet does not correspond to any locally configured segment or interface~\cite{rfc8200}.

Hereafter we report a short description of the most important behaviors starting with \textit{SR policy headend} ones. The \textit{H.Encaps} behavior encapsulates an IPv6 packet as the inner packet of an IPv6-in-IPv6 encapsulated packet. The outer IPv6 header carries the SRH header, which includes the SIDs list. The \textit{H.Encaps.L2} behavior is the same as the \textit{H.Encaps} behavior, with the difference that \textit{H.Encaps.L2} encapsulates the full received layer-2 frame rather than the IP packet (Ethernet over IPv6 encapsulation). The \textit{H.Insert} behavior inserts an SRH in the original IPv6 packet, immediately after the IPv6 header and before the transport level header. The original IPv6 header is modified, in particular the IPv6 destination address is replaced with the IPv6 address of the first segment in the segment list, while the original IPv6 destination address is carried in the SRH header as the last SID of the SIDs list.

The \textit{End} behavior represents the most basic SRv6 function among the endpoint behaviors. It replaces the IPv6 destination address of the packet with the next SID in the SIDs list. Then, it forwards the packet based by performing a lookup of the updated IPv6 Destination Address in the routing table of the node. We will refer to the lookup in the routing table as \textit{FIB lookup}, where FIB stands for Forwarding Information Base. The \textit{End.T} behavior is a variant of the \textit{End} where the FIB lookup is performed in a specific IPv6 table associated with the SID rather than in the main routing table. The \textit{End.X} behavior is another variant of the \textit{End} behavior where the packet is directly forwarded to a specified layer-3 adjacency bound to the SID rather performing any FIB lookup of the IPv6 destination address.

The \textit{End.DT6} behavior pops out SRv6 encapsulation and perform a FIB lookup with the IPv6 destination address of the exposed inner packet in a specific IPv6 table associated with the SID. It is possible to associate the default IPv6 routing table with the SID, in this case the inner IPv6 packets will be decapsulated and then forwarded on the basis of its IPv6 destination address according to the default routing of the node. The \textit{End.DX6} behavior removes the SRv6 encapsulation from the packet and forwards the resulting IPv6 packet to a specific layer-3 adjacency bound to the SID. \textit{End.DT4} and \textit{End.DX4} are respectively the IPv4 variant of \textit{End.DT6} and \textit{End.DX6}, i.e. they are used when the encapsulated packet is an IPv4 packet. The \textit{End.DX2} behavior is used for packets encapsulated at Layer 2 (e.g. with H.Encaps.L2). It pops out the SRv6 encapsulation and forwards the resulting L2 frame via an output interface associated to the SID.

Finally, other two sets of SRv6 behaviors have been defined in \cite{ietf-spring-sr-service-programming} and \cite{id-srv6-mobile-uplane} respectively for the support of Service Function Chaining of SRv6-unaware network functions and for mobile user plane functions. Some of these behaviors such as \textit{End.AD} and \textit{End.AM} are implemented in VPP and in an external Linux kernel module \cite{srext-srv6-net-prog} but not in the Linux base kernel. The details and performance evaluation of the aforementioned behaviors as well as other SRv6 \textit{endpoint} behaviors like \textit{End.B6}, \textit{End.B6.Encaps} and \textit{End.BM} have not been considered in this work and are left for future works. These \textit{endpoint} behaviors are used to steer the traffic into an SR policy by sending it to the corresponding \textit{Binding SID} (BSID). 

\subsection{SRv6 support in the Linux kernel}
\label{sec:srv6-linux}

\begin{figure}[htbp]
	\centering
	\includegraphics[width=0.4\textwidth]{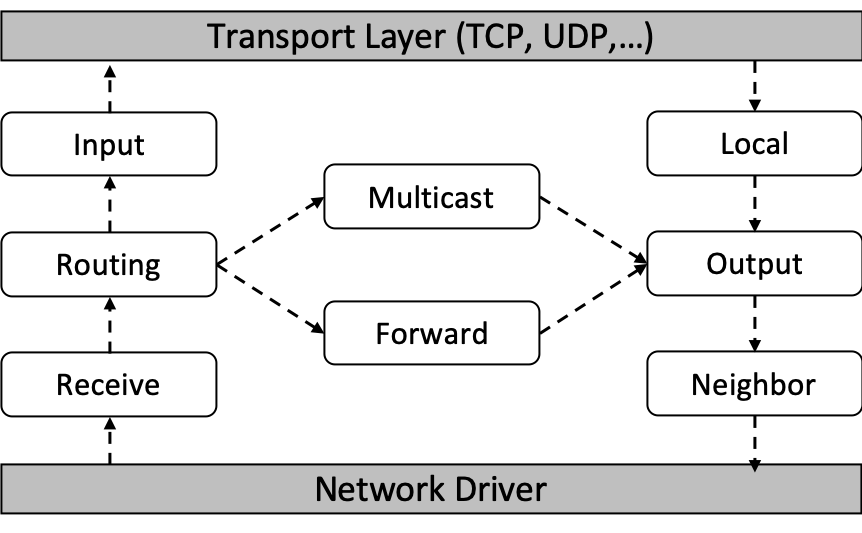}
	\vspace{-2ex}
	\caption{Linux Packet Processing Architecture}
	\label{fig:linux-pkt-process}
	\vspace{-1ex}
\end{figure}

The Linux kernel is the main component of a Linux based operating system and it is the core interface between the hardware and the user processes. The network stack in the Linux kernel can be divided into eight main subsystems; \textit{Receive}, \textit{Routing}, \textit{Input}, \textit{Forward}, \textit{Multicast}, \textit{Local}, \textit{Output} and \textit{Neighbor}. Figure \ref{fig:linux-pkt-process} shows the main subsystems of the network stack including the \textit{Network Driver}, which feeds with packets the stack and the \textit{Transport Layer} which manages local directed packets at higher levels. 

The SRv6 implementation was merged in Linux kernel 4.10~\cite{lebrun2017implementing}.
Since then, SRv6 support has become more mature in versions 4.16 and 4.18 with the addition of new features and with refinements of the implementation.
The SRH \cite{ietf-6man-segment-routing-header} is defined through a structure, named \texttt{ipv6\_sr\_hdr}. A kernel function, named \texttt{ipv6\_srh\_rcv()}, is added as a default handler for SRv6 traffic and it is called by the \textit{Receive} subsystem when an IPv6 extension header is found. The processing of received SRv6 packets is controlled through a per-interface configuration option (\texttt{sysctl}). Based on this per-interface option, the kernel may decide to either accept or drop a received SRv6 packet. If the packet is accepted, it is processed as described in~\cite{ietf-6man-segment-routing-header}: the SRH is processed, the packet IPv6 destination address is updated, then the kernel feeds the packet again in the \textit{Routing} subsystem to be forwarded based on the new destination address.

In the Linux kernel, the SRv6 behaviors are implemented as Linux lightweight tunnels (\texttt{lwtunnel}). The \texttt{lwtunnel} is an infrastructure that was introduced in the version 4.3 of the kernel to allow for scalable flow-based encapsulation such as MPLS and VXLAN. SRv6 SIDs are configured as IPv6 FIB entries into the main routing table, or into any secondary routing table~\cite{srv6-impl}. In order to support adding SIDs associated with an SRv6 behavior, the iproute2 user-space utility has been extended~\cite{iproute2}. The SRv6 capabilities were improved in the release 4.18~\cite{kernel4-18} (August 2018), to include the netfilter framework~\cite{netfilter-hacking} as well as the eBPF framework~\cite{lwn-ebpf}. 

At the time of writing, several \textit{SR policy headend} behaviors are supported in the Linux kernel, including: \textit{H.Insert}, \textit{H.Encaps}, and \textit{H.Encaps.L2}. 
As anticipated at the beginning of this section, \textit{endpoint} behaviors are classified as \textit{no-decap} and \textit{decap}. Regarding the \textit{no-decap} behaviors there is support for \textit{End}, \textit{End.T} and \textit{End.X}. 
Instead regarding the \textit{decap} functions there is support for \textit{End.DX2}, \textit{End.DT6}, \textit{End.DX6}, \textit{End.DX4}.  
Currently, the implementation of the \textit{End.DT4} behavior is missing in the Linux kernel.

\subsection{SRv6 support in VPP}
\label{sec:srv6-vpp}

Virtual Packet processor (VPP) is an open source virtual router \cite{fd-io-vpp}. It implements an high performance forwarder that can run on commodity CPUs. In addition, VPP is a very flexible and modular framework that allows the addition of new \emph{plugins} without the need to change the core kernel code. VPP often runs on top the Data Plane Development Kit (DPDk) \cite{dpdk}, which is a platform for high speed I/O operations. DPDK maps directly the network interface card (NIC) into user-space bypassing the underlying Operating System kernel.

\begin{figure}[htbp]
	\centering
	\includegraphics[width=0.487\textwidth]{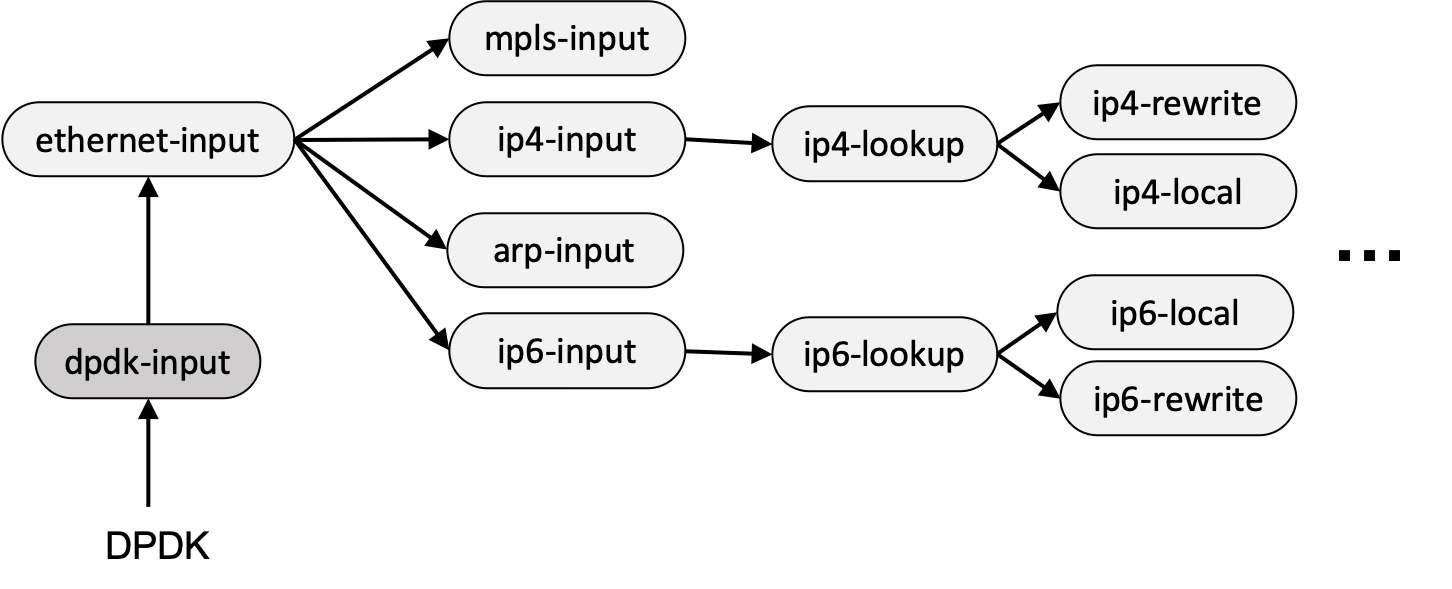}
	\vspace{-4ex}
	\caption{VPP Packet Processing Architecture}
	\label{fig:vpp-pkt-process}
	\vspace{-1ex}
\end{figure}

The packet processing architecture of VPP consists of graph nodes that are composed together. Each graph node performs one function of the processing stack such as IPv6 packets input (\emph{ip6-input}), or IPv6 FIB look-up (\emph{ip6-lookup}). The composition of the several graph nodes of VPP is decided at runtime. Figure~\ref{fig:vpp-pkt-process} shows an example of a VPP packet graph. VPP also supports batch packet processing~\cite{barach2018batched}, a technique that allows the processing of a batch of packets by one VPP graph nodes before passing them to the next node. This technique improves the packets processing performance by leveraging the CPU caches. Performance aspects of VPP are discussed in~\cite{csit-report} and \cite{barach2018batched}. 

\begin{figure*}[!t]
	\centering
	\includegraphics[trim={0cm 0.45cm 0cm 0.32cm},clip,width=0.65\textwidth]{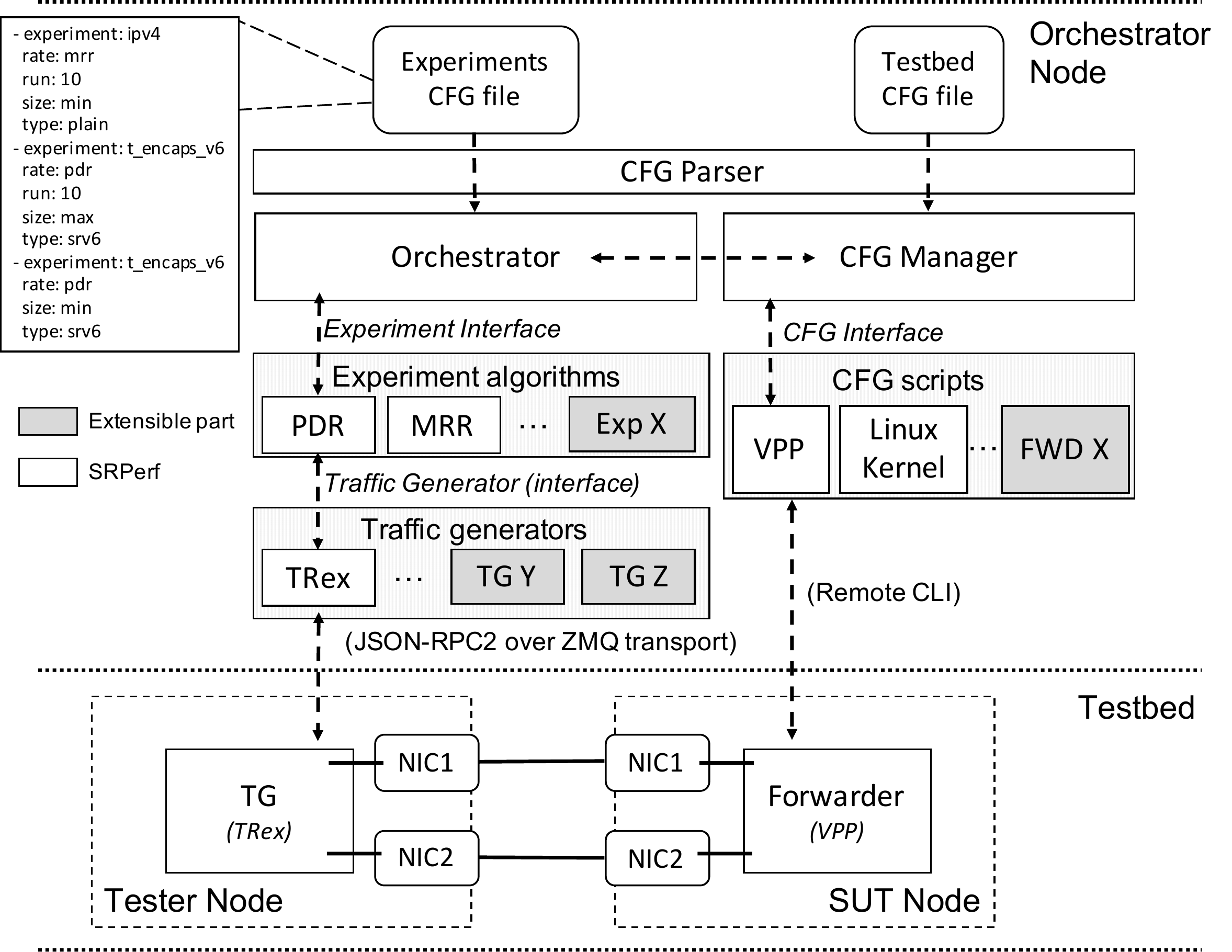}
	\vspace{-1ex}
	\caption{SRPerf architecture.}
	\label{fig:framework}
	\vspace{-2ex}
\end{figure*}

SRv6 capabilities were introduced in the 17.04 release of VPP. Most of the SRv6 \textit{endpoint} behaviors defined in~\cite{ietf-spring-srv6-network-programming} are nowadays supported (e.g. \textit{End}, \textit{End.X}, \textit{End.DX2}, \textit{End.DX4}, \textit{End.DX6}, \textit{End.DT4}, \textit{End.DX6}). These behaviors are grouped by the \textit{endpoint} function type and implemented in dedicated VPP graph nodes. For example, all the \emph{decap} functions share one single graph node, while the \textit{End} and \textit{End.X} functions are implemented in another VPP graph node. The SRv6 graph nodes perform the required SRv6 behaviors as well the IPv6 processing (e.g. decrement \textit{Hop Limit}). When an SRv6 segment is instantiated, a new IPv6 FIB entry is created for the segment address that points to the corresponding VPP graph node. An API was added to allow developers the creation of new SRv6 endpoint behaviors using the plugin framework. In this way, a developer can focus on the actual behavior implementation while the segment instantiation, listing and removal are performed by the existing SRv6 code. 

The SR policy concept was introduced to allow the \textit{SR policy headend} capabilities. Traffic can be steered into an SR policy either by sending it to the corresponding BSID or by configuring a rule, called \textit{Steering Policy}, that directs all traffic, for example transiting towards a particular IP prefix, into the policy. 
While for the \textit{SR policy headend} behaviors there is parity in the capabilities offered by Linux kernel and VPP; it is not the same for the \textit{endpoint} behaviors where VPP implementation exhibits a broader support of the SRv6 network programming model. Release 17.04 introduced the support for most of the behaviors, which included also the \textit{endpoint} behaviors bound to a policy: \textit{End.B6}, and \textit{End.B6.Encaps}. Instead, \textit{End.T} was introduced in the subsequent release (17.10). Finally, the SR proxy behaviors were introduced as VPP plugins in release 18.04~\cite{ietf-spring-sr-service-programming}.

\section{Performance evaluation framework}
\label{sec:framework}

In this section, we illustrate the proposed performance evaluation framework (SRPerf). At first, we describe the internal design and the high level architecture of SRPerf (Section \ref{sec:design}); leveraging the SRPerf modular design, we have integrated the VPP platform and the Linux kernel as \textit{Forwarder}. Section \ref{sec:methodology} elaborates on our evaluation methodology which uses the Partial Drop Rate (PDR) metric to characterize the performance of a system. Finding the PDR of a given forwarding behavior is a time consuming and error prone process, for this reason we have developed an automatic finder procedure which is described in Section \ref{sec:pdr_algorithm}. Our algorithm performs a logarithmic search in the space of the solutions and adapts to different forwarding engines and does not require manual tuning. Moreover, it is demonstrated to be efficient.

\subsection{Design and architecture}
\label{sec:design}

We designed SRPerf following the network benchmarking guidelines defined in RFC 2544~\cite{rfc2544}. As shown in Figure~\ref{fig:framework}, the architecture of SRPerf is composed of two main building blocks: the testbed and the \textit{Orchestrator}. In turn, the testbed is composed by the Tester node and the System Under Test (SUT) node. These nodes have two network interfaces cards (NIC) each and are connected back-to-back using both NICs. The Tester sends traffic towards the SUT through one NIC, which is then received back through the other one, after being forwarded by the SUT. Accordingly, the Tester can easily perform all different kinds of throughput measurements as well as round-trip delay and jitter. In our design, we chose the open source project TRex~\cite{trex-cisco} as Traffic Generator (TG) (supporting the transmission and the reception of packets in the Tester Node). As for the SUT Node, we currently support VPP and Linux kernel as \textit{Forwarder}.

Let us describe SRPerf using a top-down approach. Two configurations files (upper part of the Figure~\ref{fig:framework}) are provided as input to the \textit{Orchestrator}. The first file, \textit{Experiments CFG file}, represents the necessary input to run the experiments. In particular, it defines: i) the type of experiment (i.e. set of SRv6 behaviors to be tested, type of tests and type of algorithm); ii) the number of runs; iii) the size and type of the packets to be sent between the traffic generator and the \textit{Forwarder}. The second configuration file (\textit{Testbed CFG file}) defines the forwarding engine of the SUT and the information needed to establish a SSH connection with it. The SRPerf configuration files use the YAML~\cite{yaml} syntax, an example of configuration is reported in the upper-left part of the Figure~\ref{fig:framework}.

The \textit{Orchestrator} leverages the \textit{CFG Parser} to extract the configuration parameters and to initialize the experiment variables. The \textit{CFG Parser} is a simple python module which uses \textit{PyYAML} parser \cite{pyaml} to return python objects to the caller. The \textit{Orchestrator} is responsible for the automation of the whole evaluation process. According to the input parameters, it creates an \textit{Experiment}; specifically, the \textit{Orchestrator} uses different \textit{Experiment algorithms} for calculating the throughput. Each algorithm offers an API (the \textit{Experiment} interface in Figure~\ref{fig:framework}) through which the \textit{Orchestrator} can run an \textit{Experiment algorithm}. An example of currently supported throughput measurement algorithms is the Partial Drop Rate (PDR), described in Section~\ref{sec:methodology}. Moreover, the \textit{Orchestrator} provides a mapping between the forwarding behaviors to be tested and the type of traffic required to test each behavior. For example, to test the \textit{End} behavior, it is necessary to use an SRv6 packet with an SRH containing a SID list of at least two SIDs and the active SID must not be the last SID - the type of packet to be replayed during the experiments has to be passed to the Experiment algorithm instance.

The \textit{Orchestrator} controls the TG (deployed in the Tester node) through the high level abstraction provided by the \textit{TG Driver}, which translates the calls coming from the other modules in commands to be executed on the TG. Each driver is a python wrapper that can speak native python APIs or use any other transport mechanism supported by the language. For example, the TRex driver includes the python client of the TRex automation API \cite{trex-client-api} that uses as transport mechanism JSON-RPC2 \cite{jsonrpc} over ZMQ \cite{zmq}. The \textit{Orchestrator} can be deployed on the same node of the TG or in a remote node. 

The \textit{CFG Manager} controls the forwarding engine in the SUT. It is responsible for enforcing the required configuration in the \textit{Forwarder}. The \textit{Orchestrator} provides the mapping between the forwarding behaviors to be tested and the required configuration of a given forwarding engine. Hence, the \textit{Orchestrator} is able to properly instruct the \textit{CFG Manager}. For each forwarding engine, we implement a \textit{CFG script} which provides the \textit{CFG Manager} with the means to enforce a required configuration. In particular, a \textit{CFG script} is a bash script defining a configuration procedure for each behavior to be tested. The configuration is applied using the Command Line Interface (CLI) exposed by the forwarder. For example, to test the \textit{End} behavior in the Linux kernel, we implement a bash procedure called \textit{end}. In this procedure, we leverage the \textit{iproute} utility to configure the forwarding engine in the SUT with two FIB entries: 1) an SRv6 SID with the \textit{End} behavior; 2) a plain IPv6 FIB entry to forward the packet once the \textit{End} function has been performed. The configuration can be as simple as adding a FIB entry to forward the received packets back to the Tester, but also being a more complex configuration that manipulates the incoming packets before forwarding them back to the Tester. The \textit{CFG Manager} first pushes the \textit{CFG scripts} in the SUT and then applies a given configuration running commands over an SSH connection.

The SRPerf implementation is open source and available at~\cite{srperf}. SRPerf is mostly written in python, and provides a set of tools to facilitate the deployments of the experiments: it offers an API for the automatic generation of the configuration files. Moreover, it provides different installation scripts to setup a performance evaluation experiment using SRPerf on any commodity hardware. These scripts include \textit{TG} installation and initial configuration, \textit{Forwarder} installation and initial configuration. The framework is modular and can be expanded in different directions: it can be extended to support new traffic generators by simply creating a new driver for each. A new forwarding behavior can be added by updating the \textit{CFG Manager} with the configuration required for such behavior. New algorithms for calculating throughput and delay can be developed and plugged into the \textit{Orchestrator}. It can support different \textit{Forwarders} in the SUT, which only requires the \textit{CFG manager} to be updated to recognize them and to implement the related \textit{CFG script}. In this work we have first considered the Linux kernel networking as Forwarder and then, leveraging the framework described above, we added the support for VPP software router.

\begin{figure}[t]
	\centering
	\includegraphics[trim={0cm 0.25cm 0cm 0.25cm},clip,width=0.487\textwidth]{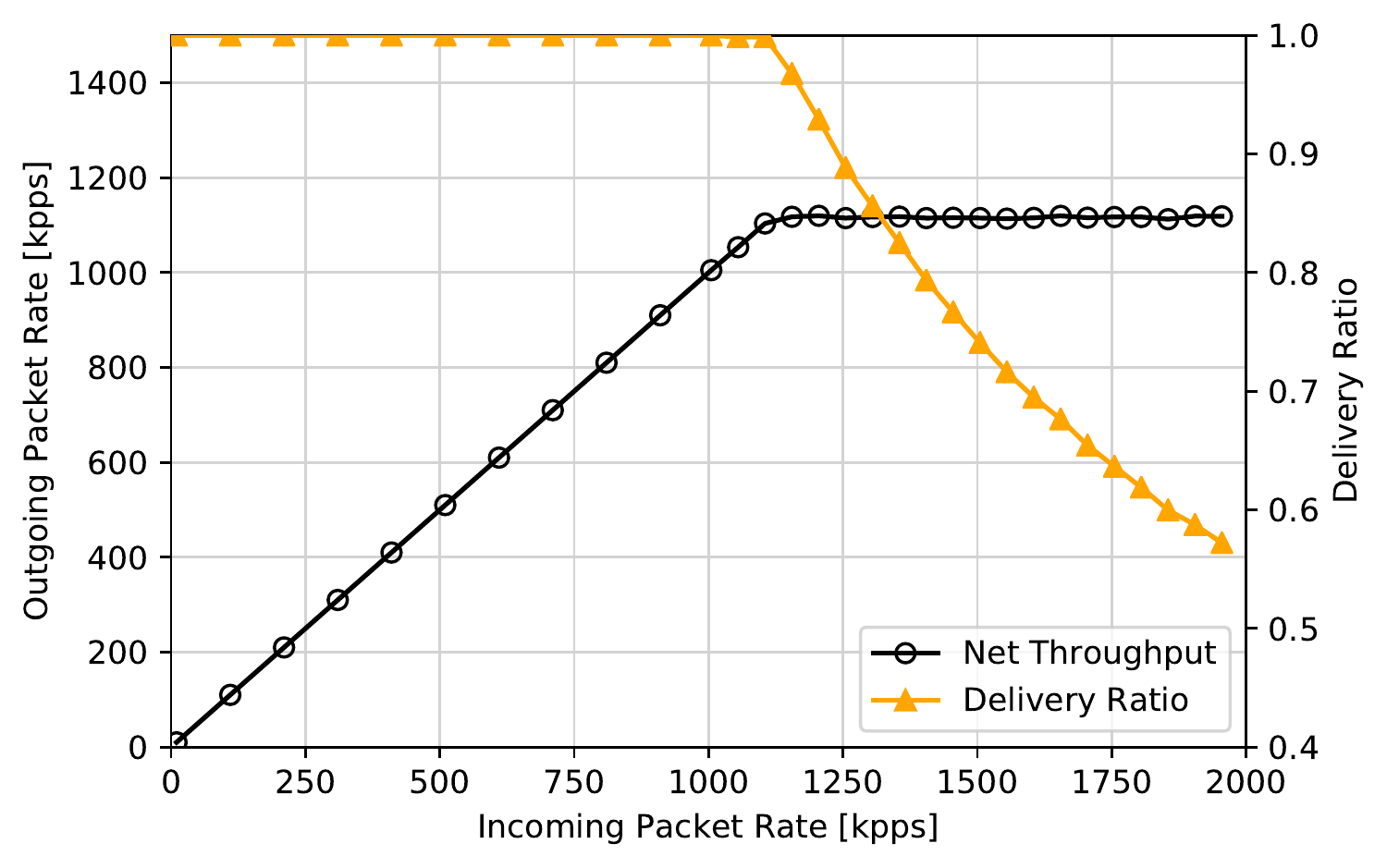}
    \caption{Throughput of plain IPv6 forwarding}
    \label{fig:no-drop-region}
    \vspace{-2ex}
\end{figure}

\subsection{Evaluation methodology}
\label{sec:methodology}

RFC 1242~\cite{rfc1242} defines the \textit{throughput} as the maximum rate at which all received packets are forwarded by the device. RFC 1242 suggests that this should be used as a standard measure to compare performance of network devices from different vendors. Throughput can be reported in number of bits per second (bps) as well as number of packet per second (pps). FD.io CSIT Report \cite{csit-report} defines \textit{No-Drop Rate} (NDR) and \textit{Partial Drop Rate} (PDR). NDR is the highest forwarding rate achieved without dropping packets, so it corresponds to the \textit{throughput} defined by RFC 1242. PDR is the highest received rate supported without dropping traffic more than a pre-defined loss ratio threshold \cite{opnfv-nfvbench}. We use the notation PDR@X\%, where X represents the loss ratio threshold. For example, we can evaluate PDR@0.1\%, PDR@0.5\%, PDR@1\%. NDR can be described as PDR@0\%, i.e. PDR with a loss threshold of 0\%. Considering that the term \textit{throughput} can be used with wider meanings, the terminology defined in \cite{csit-report} (e.g. No-Drop Rate) is clearer and it will be used hereafter. Hence, we will use \textit{throughput} to refer in general to the output forwarding rate of a device. 
In this work, we will consider only the PDR since it is more generic than the NDR and we will express all rates in packets per second. 

Finding the PDR requires the scanning of a broad range of possible traffic rates. In order to explain the process, let us consider the plain IPv6 forwarding in the Linux kernel. Figure~\ref{fig:no-drop-region} plots the throughput (i.e. the output forwarding rate) and the \textit{Delivery Ratio} (DR) versus the input rate, defined and evaluated as follows. We generate traffic at a given packet rate $PS$ [kpps] for a duration $D$ [s] (usually $D=10 s$ in our experiments). Let the number of packets generated by the TG node and incoming to the SUT in an interval of duration $D$ be $P_{IN}$ (Packets INcoming in the SUT). We define the number of packets transmitted by the SUT (and received by the TG) as $P_{OUT}$ (Packets OUTgoing from the SUT). The throughput $T$ is $P_{OUT}/D$ [kpps]. We define the DR as $P_{OUT}/P_{IN}=P_{OUT}/(PS*D)=T/PS$. 
Hence, the DR is the ratio between the input and the output packet rates of a device for a given forwarding behavior under analysis. It is 100\% for all incoming data rates less than the device No-Drop Rate. Initially, the throughput increases linearly with the increase in the incoming rate. This region is often referred to as \textit{no drop region}, i.e. where the DR is always 100\%. If the forwarding process is CPU-limited, the CPU usage at the SUT node increases with the increase of incoming traffic rate (i.e. the sending rate of the Tester). Ideally, the SUT node should be able to forward all received packets until it becomes 100\% CPU saturated. On the other hand, in our experiments with the Linux based SUT we measured very small but not negligible packet loss ratio in a region where we have an (almost) linear increase of the \textit{Throughput}. Therefore, it is better to consider the \textit{Partial Drop Rate} and we used 0.5\% as threshold. The PDR@0.5\% is the highest incoming rate at which the \textit{Delivery Ratio} is at least 0.995. The usefulness of the PDR is that it allows to characterize a given configuration of the SUT with a single scalar value, instead of considering the full relation between incoming rate 
and throughput shown in Figure~\ref{fig:no-drop-region}. The procedure for finding the PDR for a given loss threshold is described in the Section \ref{sec:pdr_algorithm} hereafter. 


Finally, we define LPR (Line Packet Rate) as the maximum packet rate that can be achieved considering the line bit rate R and the size of the packets used during an experiment: 

\begin{equation}
LPR = R / [8*(FrameSize + Overhead)]
\label{eq:linerate}
\end{equation}
\vspace{-2ex}

Where \textit{R} [bps] is the line bit rate (e.g. $10*10^{9}$ for 10GbE, \textit{FrameSize} is the frame size in bytes at Ethernet level (including the 14 bytes of Ethernet header), the \textit{Overhead} for the Ethernet frames is 24 bytes (4 for CRC, 8 for preamble/SFD and 12 for the inter frame gap). Clearly, the PDR rate is upper limited by the line packet rate LPR. 

\subsection{PDR finder algorithm}
\label{sec:pdr_algorithm}

Estimating the PDR of a given forwarding behavior is a time consuming process, since it requires the scanning of a broad range of possible traffic rates. In order to automate the PDR finding process, we have designed and developed the PDR finder algorithm. It scans a range of traffic rates with the objective of estimating the PDR value. Alg.~\ref{alg:pdr_finder} reports the pseudo code of the PDR finder algorithm. It performs a logarithmic search in the space of possible solutions which is upper limited by the line packet rate LPR of the NICs. It returns an interval $[a,b]$ of traffic rates estimating the PDR value with a given accuracy $\epsilon$. The accuracy is configurable to tune the algorithm precision. The algorithm starts to decrease the size of the search interval until it becomes less than the desired accuracy (line \ref{lst:return-condition}). At each iteration (loop starting at line \ref{lst:loop}) the DR is evaluated for the middle point of the search interval, which is used as the packet generation rate in the TG node. If the measured Delivery Ratio is less than the loss threshold, the upper bound of the search interval is set to the current rate. Otherwise, the lower bound of the search interval is set with the current rate. In this way, the size of the search interval is halved. This process is iterated until the exit condition is triggered: the algorithm terminates when the difference between $a$ and $b$ is less than or equal to $\epsilon$. The algorithm takes as input the initial values of the search interval (min, max) and the required accuracy, all expressed as percentage of the line packet rate. 

\begin{algorithm}[htpb]
\caption{PDR finder algorithm}
\label{alg:pdr_finder}
\begin{algorithmic}[1]
\REQUIRE $linePacketRate, lossThreshold,$ $min, max, accuracy$
\STATE $lowBound \leftarrow linePacketRate*min/100$ 
\STATE $upBound \leftarrow linePacketRate*max/100$ 
\STATE $\epsilon \leftarrow linePacketRate*accuracy/100$ 
\LOOP \label{lst:loop}
 \STATE // The algorithm terminates when the size of the searching window is less than the threshold $\epsilon$
 \IF{$\left| upBound-lowBound \right| \leq \epsilon$} \label{lst:return-condition}
 \RETURN $\left[ lowBound, upBound \right]$
 \ENDIF
 \STATE // Evaluate the DR for the window middle point
 \STATE $txRate \leftarrow \dfrac{lowBound+upBound}{2}$
 \STATE $rxRate \leftarrow runExperiment(txRate)$ \label{lst:runExp}
 \STATE $deliveryRatio \leftarrow \dfrac{rxRate}{txRate}$ \label{lst:delRatio}
 \STATE // Halve the size of the searching window
 \IF{$deliveryRatio < (1-lossThreshold)$}
 \STATE $upBound \leftarrow txRate$
 \ELSE
 \STATE $lowBound \leftarrow txRate$
 \ENDIF
 \ENDLOOP
\end{algorithmic}
\end{algorithm}
\vspace{-2ex}

A caveat is needed with respect to the conceptual algorithm described in Alg.~\ref{alg:pdr_finder}. The Delivery Ratio evaluated in line \ref{lst:delRatio} is based on one experiment run at a given \textit{txRate} (the rate of packets sent by the Traffic Generator) which evaluates an \textit{rxRate} (the rate of packets forwarded by the SUT and received by the TG). The variability of the experiment result (i.e. the number of forwarded packets) must be carefully considered, because it affects the evaluation of the delivery ratio. This means that the experiment mentioned in line \ref{lst:runExp} should be repeated multiple times, the variance of the result should be evaluated and the result should be accepted when the variance is below a given threshold. Note that, in order to increase the overall efficiency of the PDF finder algorithm, this accurate check is needed when the Delivery Ratio is close to the threshold value, while it is not needed when the Delivery Ratio is 1 and when the Delivery Ratio is much lower than the threshold. 

Finally, we validate the PDR finder procedure to make sure that the estimated PDR value is stable. In particular, to calculate a PDR value we run a number of overall repetitions (e.g. 10) to evaluate the standard deviation of the $P_{OUT}$ across these repetitions.

\section{Performance evaluation of the SRv6 software implementations}
\label{sec:performance}

In this section, we present an evaluation of two SRv6 software implementations, namely Linux kernel and VPP software router. The rationale for this evaluation is to provide an indication on the scalability of the SRv6 implementations over a set of experiments. It is not our purposes to make a direct comparison of the dataplane forwarding performance between the two implementations, as they are internally very different, making their comparison pointless. Section \ref{sec:testbed} illustrates the testbed and the parameters of the experiments. We report in Section \ref{sec:linux_exp} the experiment results of the Linux kernel forwarding. During the code analysis of the Linux kernel implementation, we discovered the \textit{End.DT4} was missing, Section \ref{sec:end_dt4} illustrates how we have used SRPerf to benchmark the experimental implementation of the \textit{End.DT4} we have realized. Instead, Section \ref{sec:end_low} shows how we have leveraged SRPerf to solve the performance issues we found in some \textit{endpoint} behaviors. Finally, Section \ref{sec:vpp_exp} reports the experiments results of VPP.

\subsection{Testbed and parameters of the experiments}
\label{sec:testbed}

Our testbed, illustrated in the bottom part of Figure~\ref{fig:framework}, has been deployed on CloudLab~\cite{cloudlab}. The testbed nodes (\textit{Tester} and \textit{SUT}) are powered by a bare metal server equipped with an Intel Xeon E5-2630 v3 processor with 16 cores 
clocked at 2.40GHz and 128 GB of RAM. Each bare metal server has two Intel 82599ES 10-Gigabit network interface cards to provide back-to-back connectivity between the testbed nodes. The \textit{Tester} is running the TRex~\cite{trex-cisco} traffic generator and has the TRex python automation libraries installed. The \textit{SUT} machine is running Linux kernel 5.2 \textit{net-next} (upstream branch of the Linux kernel). It has the 5.x release of the \textit{iproute2}~\cite{iproute2} installed, which provides the means to program the SRv6 behaviors. In addition, \textit{ethtool} (release 5.x) is installed to provide the means to configure the NIC hardware capabilities such as offloading~\cite{ethtool}. Regarding VPP, we have been using the release 19.04. 

\begin{figure}[htbp]
	\centering
	\includegraphics[width=0.45\textwidth]{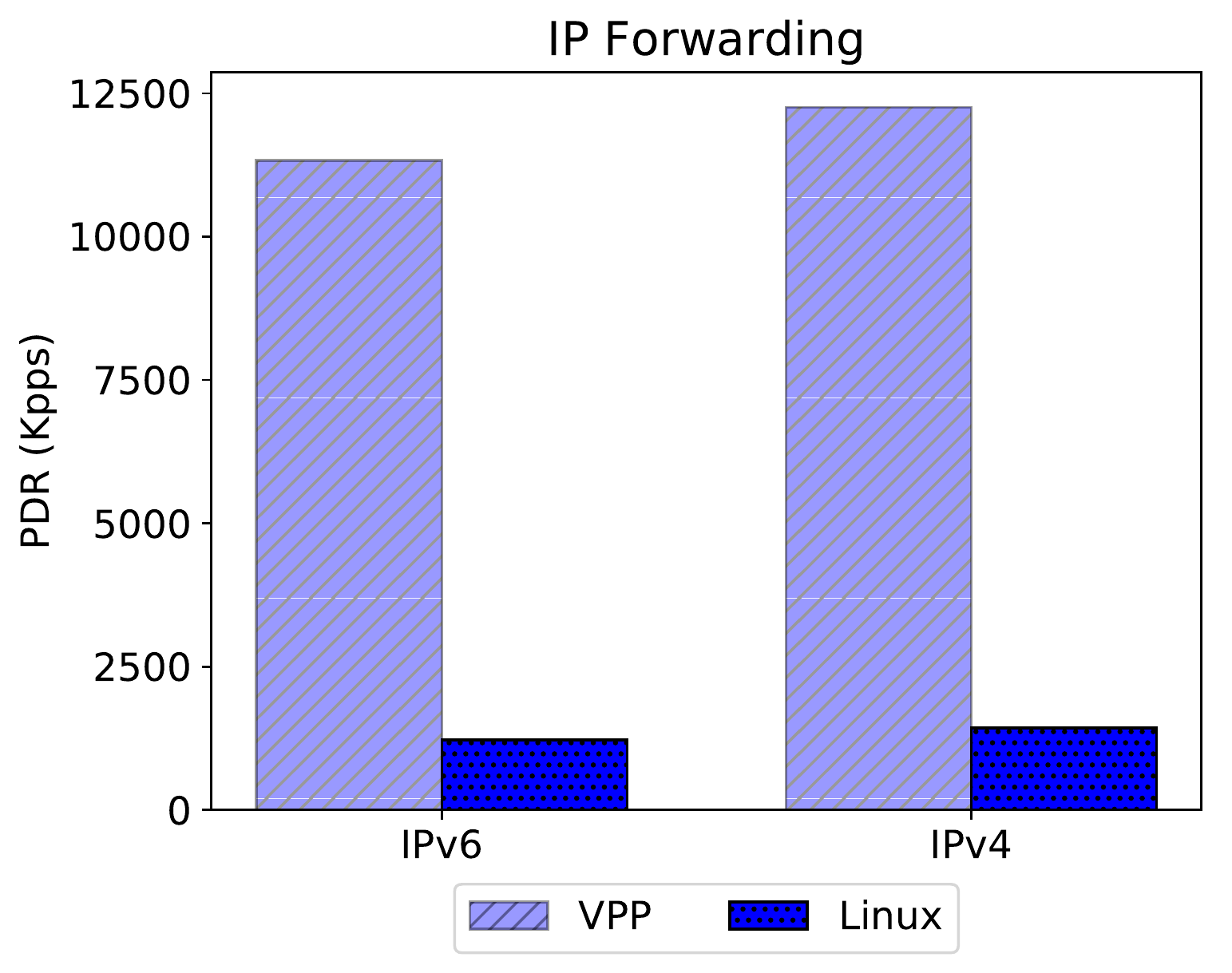}
	\vspace{-2ex}
    \caption{Plain IP forwarding} 
    \label{fig:ip}
	\vspace{-1ex}
\end{figure}

Before discussing the results of the experiments, let us describe the methodology we have used to perform the experiments and some tuning parameters. We configured a single CPU core in our SUT for the processing and forwarding of the incoming packets. These single-core measurements provide the base performance for a given behavior, obviously when multiple cores are allowed to process packets the performance scales up. Regarding the Linux kernel, in order to force the single CPU core processing of all received traffic, we rely on the Receive-Side Scaling (RSS) and SMP IRQ affinity features. Moreover, in order to get the base performance independent of the NIC hardware capabilities, we disabled all the NIC hardware offloading capabilities such as Large Receive Offload (LRO), Generic Receive Offload (GRO), Generic Segmentation Offload (GSO), and all checksum offloading features. Finally, we disabled the hyper-threading feature of the SUT node from the BIOS settings. Further details about the tuning of these features are reported in our previous work \cite{ahmedperformance}. Similar configurations have been performed for VPP. In particular, as VPP is a user space router we had just to customize the startup configurations to use one CPU core and to disable all the DPDK offloading features. 
We configured the \textit{TUNSRC} for the SRv6 \textit{policy headend} behaviors doing encapsulation. The latter allows to configure the IPv6 source address of the IPv6 outer header. The \textit{TUNSRC} has to be configured otherwise the Linux kernel will try to get the address from the interface which will cause a performance drop in the performance of the encaps behavior.

We classified the forwarding behaviors into three classes as follows: i) \textit{SR policy headend} behaviors; ii) \textit{endpoint} behaviors with no decapsulation (no-decap); iii) \textit{endpoint} behaviors with decapsulation (decap). The \textit{SR policy headend} behaviors receive non-SRv6 traffic and adds the SRH header, either inserting it in an IPv6 packet (\textit{H.Insert}) or encapsulating the received packet in an outer IPv6 packet with the SRH header (e.g. \textit{H.Encap}). The \textit{decap} behaviors are required to remove the SRv6 encapsulation from packets before forwarding them. Conversely, the \textit{no-decap} behaviors forward SRv6 packets without removing the SRv6 encapsulation from packets. For the SRv6 \textit{policy headend} behaviors experiments, we use an IPv6 packet of size 64 bytes. For all the SRv6 \textit{endpoint} behaviors, we use an inner IPv6 packet of size 64 bytes plus the SRv6 encapsulation (80 bytes, i.e. 40 bytes of outer IPv6 header and 40 bytes of SRH with two SIDs). 

We use the PDR metric described in Section~\ref{sec:methodology} (in particular we consider PDR@0.5\%). The trail period in our experiments is 10 seconds. We use the bar plots to represent our results, where each bar plot represents the average of 10 PDR values. The reported PDR value is the average of 10 repetitions. Table \ref{tab:pe1_linux}, \ref{tab:pe2_linux}, \ref{tab:pe1_vpp} and \ref{tab:pe2_vpp} respectively report the average, the Coefficient of Variation (CV) and the 95\% Confidence Interval ($CI_{95}$) of the PDR (measured in kpps) for each analyzed forwarding behavior. Note that as discussed in the previous section, the PDR rate is upper limited by the line packet rate which depends on the size of the packets used during the experiment. For a 10GbE interface and an IP packet of 64 bytes, the line packet rate is $\approx$12255 kpps. 

In a preliminary experiment, we evaluated the performance of the plain IP forwarding for both Linux kernel and VPP, over a 10GbE interface. Figure \ref{fig:ip} reports the results for an IP packet length of 64 bytes. In this test, we can state that VPP is able to forward the IPv4 packets at the line packet rate ($\approx$12252), while for IPv6 traffic where the forwarding rate is lower ($\approx$11327) than the line packet rate. The performance of the Linux kernel is lower, we measured $\approx$1221 kpps and $\approx$1430 kpps respectively for IPv6 and IPv4. Similarly to VPP, the forwarding of IPv4 traffic results to be more performant.

\begin{figure*}[!t]
	\centering
	\includegraphics[trim={0cm 0.4cm 0cm 0.3cm},clip,width=0.75\textwidth]{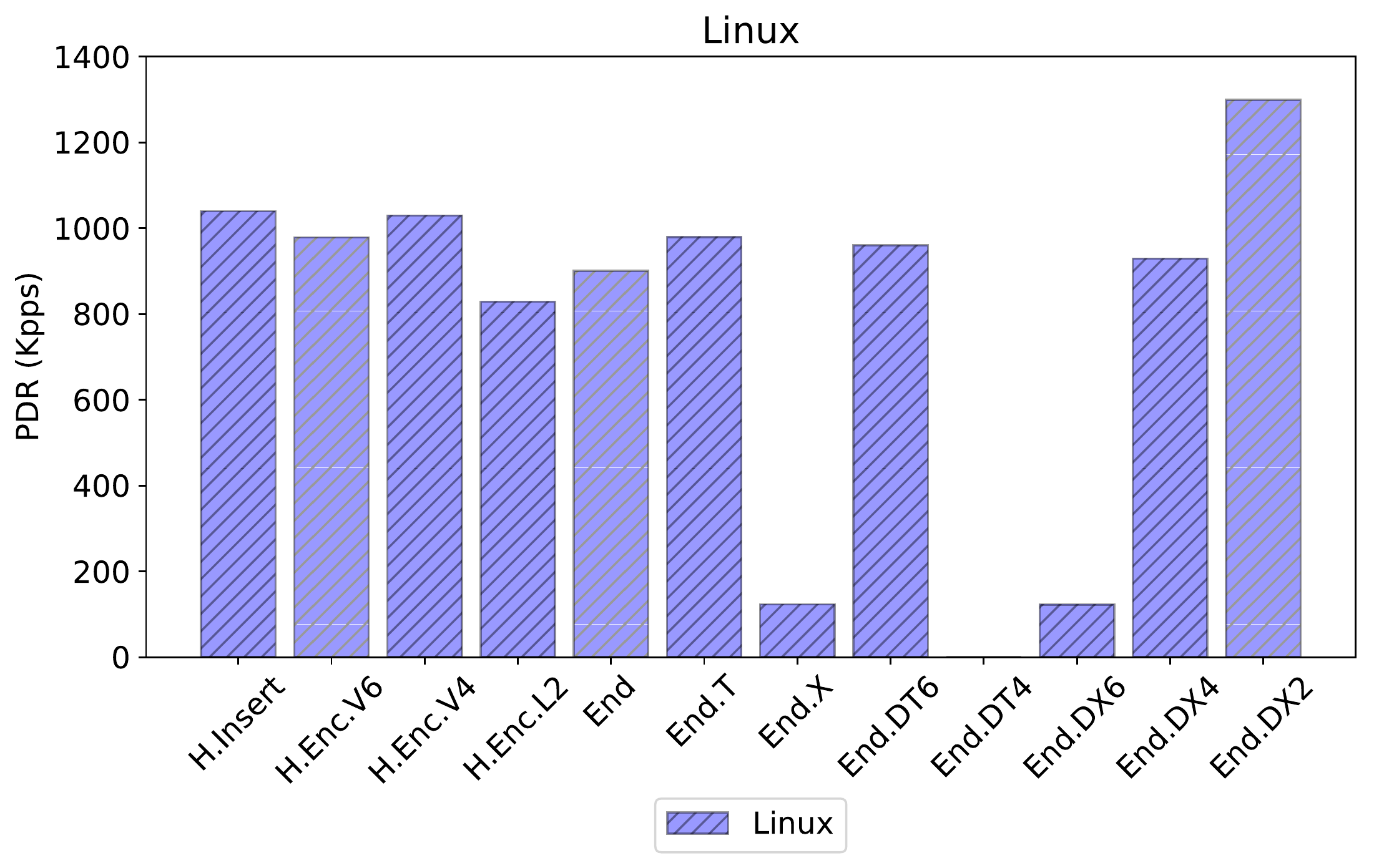}
	\vspace{-1ex}
    \caption{Linux kernel results}
    \label{fig:linux}
	\vspace{-1ex}
\end{figure*}

\begin{table*}[!ht]
\scalebox{1}{
    \begin{tabular}{|c|c|c|c|c|c|c|}
      \hline
        & IPv6 & IPv4 & H.Insert & H.Enc.V6 & H.Enc.V4 & H.Enc.L2 \\
      \hline
      Mean      & 1221.06 & 1430.38 & 1039.29 & 978.133 & 1029.35 & 828.891 \\
      \hline
      CV        & 0.023\% & 0.009\% & 0.016\% & 0.009\% & 0.081\% & 0.019\% \\
      \hline
      $CI_{95}$ & 0.014\% & 0.006\% & 0.01\% & 0.006\% & 0.051\% & 0.012\% \\
      \hline
    \end{tabular}
    }
\centering 
\vspace{-1ex}
\caption{Plain IP forwarding and \textit{SR policy headend} in Linux kernel. Mean, CV and $CI_{95}$}
\label{tab:pe1_linux}
\end{table*}

\begin{table*}[!ht]
\scalebox{1}{
    \begin{tabular}{|c|c|c|c|c|c|c|c|c|}
      \hline
       & End & End.T & End.X & End.DT6 & End.DT4 & End.DX6 & End.DX4 & End.DX2 \\
      \hline
      Mean      & 900.52 & 979.253 & 123.133 & 960.061 & N/A & 122.761 & 929.022 & 1299.15 \\
      \hline
      CV        & 0.059\% & 0.133\% & 1.12\% & 0.078\% & N/A & 0.572\% & 0.001\% & 0.0\% \\
      \hline
      $CI_{95}$ & 0.037\% & 0.084\% & 0.71\% & 0.049\% & N/A & 0.362\% & 0.0\% & 0.0\% \\
      \hline
    \end{tabular}
    }
\centering 
\vspace{-1ex}
\caption{SRv6 \textit{endpoint} behaviors in Linux kernel. Mean, CV and $CI_{95}$}
\label{tab:pe2_linux}
\vspace{-2ex}
\end{table*}

\subsection{Linux kernel}
\label{sec:linux_exp}

We start evaluating the performance of the \textit{SR policy headend} behaviors: \textit{H.Insert}, \textit{H.Encaps} (considering IPv4 and IPv6 traffic), and \textit{H.Encaps.L2}. The results are shown in Figure~\ref{fig:linux} for Linux kernel. The \textit{H.Insert} shows a better forwarding rate of $\approx$1039 kpps when compared to $\approx$978 kpps and 1029 kpps of \textit{H.Encaps.V6} and \textit{H.Encaps.V4}. For the \textit{H.Encaps.L2} behavior, the SUT node is able to forward $\approx$828 kpps. The performance of \textit{H.Insert} behavior is slightly better compared to \textit{H.Encaps} since the former needs to push only an SRH while the latter needs to push an outer IPv6 header along with the SRH. As expected, the encap of IPv4 traffic performs better of its IPv6 counterpart. In general, the \textit{SR policy headend} behaviors have shown very stable performance as witnessed by the low values for the CV shown in Table \ref{tab:pe1_linux}. 

Regarding the \textit{no-decap} SRv6 \textit{endpoint} behaviors, we evaluated the performance of the \textit{End}, \textit{End.T}, and \textit{End.X} behaviors. In case of the \textit{End} behavior the SUT node is able to forward $\approx$900 kpps. The \textit{End.T} performs better than the \textit{End} since the routing table used for the lookup is defined by the control plane, hence the kernel saves the cost of performing IP rules lookup that are executed in case of the \textit{End} behavior. The \textit{End.T} forwarding performance is $\approx$979 kpps. As regards \textit{End.X}, we found very poor performance. Forwarding rate is $\approx$123 kpps. In Section \ref{sec:end_low} we provide more insights about this low performance and we show how we have fixed this issue and achieved performance results in line with the other behaviors.

Our last set of experiments compares the performance of the SRv6 \textit{decap} behaviors. The \textit{End.DX2} behavior has a throughput of $\approx$1299 kpps which is better than the other behaviors. The reason why \textit{End.DX2} is performing better than IPv6 forwarding for example is that the kernel does not need to perform Layer-3 lookup once the packet has been decapsulated. Indeed, it pushes the packet directly into the transmit queue of the interface towards the next-hop. Instead, \textit{End.DX4} exhibits a rate $\approx$929 kpps. As for \textit{endpoint} behaviors with lookup on a specific table, namely \textit{End.DT6}, we have a performance of $\approx$960 kpps.

In general, the performance of \textit{endpoint} behaviors (both \textit{decap} and \textit{no-decap}) is less stable with respect to the \textit{SR policy headend} behaviors: the values of (CV) and $CI_{95}$ are higher as shown in Table \ref{tab:pe2_linux}. Moreover, we have found some specific problems: firstly \textit{End.DT4} is missing in the Linux kernel and then the PDR of \textit{End.DX6} and \textit{End.X} is $\approx$122 kpps, which is much lower compared to the PDR of the other behaviors. As for \textit{End.DT4}, Section \ref{sec:end_dt4} illustrates how we have implemented \textit{End.DT4} and used SRPerf to benchmark the code under development. As for \textit{End.DX6} and \textit{End.X}, in Section \ref{sec:end_low} we show how we have fixed the problems identified by the SRPerf tool.

\subsection{Adding End.DT4 behavior to the Linux kernel}
\label{sec:end_dt4}


In the context of the ROSE project \cite{rose} (which aims to realize an open SRv6 ecosystem), we have realized an implementation of the \textit{End.DT4} behavior and we have leveraged SRPerf to assess the correctness of the patch. The implementation is publicly available and we plan to submit the code to the Linux mainline; the source code of the \textit{End.DT4} behavior is available at \cite{fix_repo}. 

Firstly, we have verified that the functionality was correctly implemented. Then, we needed to stress our implementation and assess its efficiency. SRPerf is a valuable tool for both tasks. Indeed, it can be used to stress the machine for a long time pushing packets at line-rate (to verify for example that there are no memory leaks) but also to evaluate how the new behavior affects existing code. 

Thanks to SRPerf we were able to discover that the functionality was realized correctly and no memory leaks were observed in the long runs. Our first implementation was not efficient as we expected. Indeed, we were able to obtain a Partial Drop Rate of only $\approx$600 kpps while the IPv6 counterpart (i.e. \textit{End.DT6}) was able to deliver a PDR of $\approx$950 kpps. At this point, we decided to go through a different approach which required more coding and to expose some functionality of the Linux routing to SRv6. With this second attempt we were able to fill the gap we found in the performance and obtain a PDR of $\approx$980 kpps which is aligned with the expected performance. 

\subsection{Fixing a forwarding behavior in the Linux kernel}
\label{sec:end_low}

During our first evaluation, we found that the \textit{End.X} and \textit{End.DX6} behaviors exhibited poor performances and less stability 
with respect of the other SRv6 \textit{endpoint} behaviors. 
The \textit{End.X} and \textit{End.DX6} behaviors perform the cross-connection to a layer 3 adjacency which is selected when the behavior is configured. The \textit{End.DX6} operates on an IPv6 packets encapsulated in an outer SRv6 packet, by performing first the decapsulation and then the cross-connection. The \textit{End.X} operates on an SRv6 packet performing the cross-connection without decapsulation. The current implementation of these two behaviors in Linux is not fully compliant with their specifications in the SRv6 network programming document~\cite{ietf-spring-srv6-network-programming}. The IETF document specifies that SRv6 cross-connect behaviors are used to cross connect packet to the next hop through a specific interface. Instead, the current implementation uses an IPv6 next-hop provided when the behavior is configured, 
to perform a route-lookup and find the outgoing interface. This route lookup is executed on each packet to be forwarded with these behaviors. To make things worse, for an implementation problem the routing subsystem is not able to cache the result of the route-lookup as it normally happens for IPv6 packet forwarding. Therefore the PDR rate achieved by these behavior is less than the 20\% of the PDR of other behaviors.
\begin{figure*}[!t]
	\centering
	\includegraphics[trim={0cm 0.4cm 0cm 0.3cm},clip,width=0.8\textwidth]{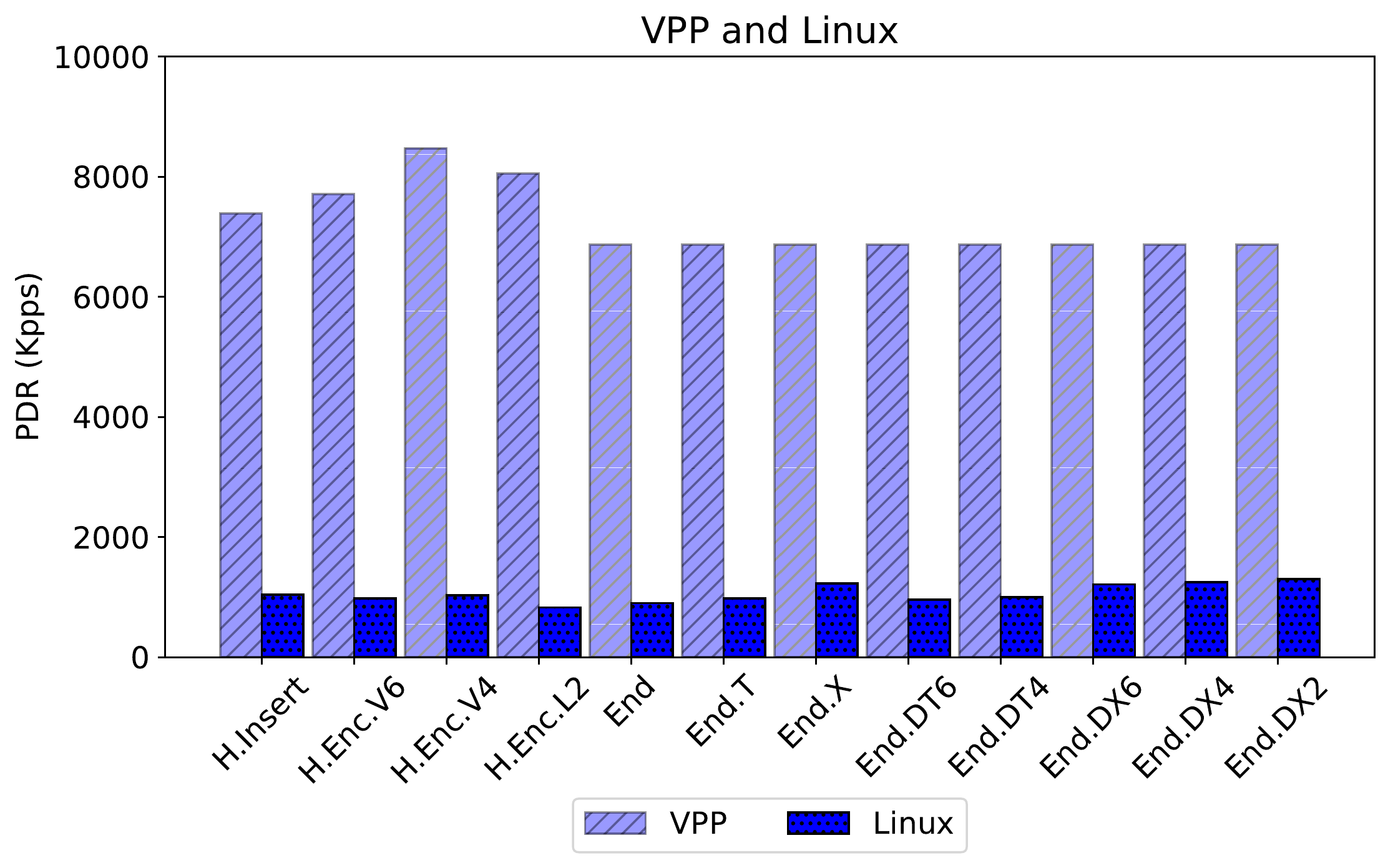}
	\vspace{-1ex}
    \caption{VPP and Linux kernel results}
    \label{fig:vpp_linux}
	\vspace{-1ex}
\end{figure*}

\begin{table*}[!ht]
\scalebox{1}{
    \begin{tabular}{|c|c|c|c|c|c|c|}
      \hline
        & IPv6 & IPv4 & T.Insert & T.Encaps.V6 & T.Encaps.V4 & T.Encaps.L2 \\
      \hline
      Mean      & 11327.5 & 12252.63 & 7387.16 & 7709.83 & 8471.8 & 8052.85 \\
      \hline
      CV        & 0.005\% & 0.0\% & 0.002\% & 0.022\% & 0.02\% & 0.0\% \\
      \hline
      $CI_{95}$ & 0.003\% & 0.0\% & 0.001\% & 0.014\% & 0.013\% & 0.0\% \\
      \hline
    \end{tabular}
    }
\centering 
\vspace{-1ex}
\caption{Plain IP forwarding and \textit{SR policy headend} behaviors in VPP. Mean, CV and $CI_{95}$}
\label{tab:pe1_vpp}
\end{table*}

\begin{table*}[!ht]
\scalebox{1}{
    \begin{tabular}{|c|c|c|c|c|c|c|c|c|}
      \hline
       & End & End.T & End.X & End.DT6 & End.DT4 & End.DX6 & End.DX4 & End.DX2 \\
      \hline
      Mean      & 6867.59 & 6867.59 & 6867.59 & 6867.59 & 6867.59 & 6867.59 & 6867.59 & 6867.59 \\
      \hline
      CV        & 0.0\% & 0.0\% & 0.0\% & 0.0\% & 0.0\% & 0.0\% & 0.0\% & 0.0\% \\
      \hline
      $CI_{95}$ & 0.0\% & 0.0\% & 0.0\% & 0.0\% & 0.0\% & 0.0\% & 0.0\% & 0.0\% \\
      \hline
    \end{tabular}
    }
\centering 
\vspace{-1ex}
\caption{SRv6 \textit{endpoint} behaviors in VPP. Mean, CV and $CI_{95}$}
\label{tab:pe2_vpp}
\vspace{-2ex}
\end{table*}

To fix the poor performance of these two cross-connect behaviors we re-designed their implementation, forwarding the packets based on a specified outgoing interface instead of using the next-hop to perform a route lookup per each packet. We implemented a new kernel function, named \texttt{seg6\_xcon6}, which is called by the \textit{End.X} and \textit{End.DX6} to cross-connect the IPv6 packet to a given interface. We have also addressed the \textit{End.DX4} behavior. Its implementation was also based on the next-hop definition and then on a route lookup, but it was not suffering of the issues discussed for \textit{End.X} and \textit{End.DX6} (because of the use of IPv4 routing instead of IPv6 routing). Nevertheless, we have re-implemented allowing to specify an outgoing interface instead of an IPv4 next-hop and obtained a significant gain in performance. 
The source code of the fixed behaviors is available at \cite{fix_repo}. 
We verified the goodness and the stability of our patch through SRPerf and we were able to obtain $\approx$1245 kpps, $\approx$1210 kpps and $\approx$1231 kpps respectively for \textit{End.DX4}, \textit{End.DX6} and \textit{End.X}. The original PDR for \textit{End.DX4} was $\approx$929 kpps. Also in this case, it is possible to appreciate the better performance of the IPv4 forwarding. The final results of the Linux kernel forwarding performance are reported in Figure~\ref{fig:vpp_linux}.

\subsection{VPP software router}
\label{sec:vpp_exp}

We have repeated the experiments performed on the Linux kernel SRv6 implementation using the VPP software router. 

We considered the SRv6 \textit{policy headend} behaviors, whose results are reported in Table \ref{tab:pe1_vpp} along with the plain IPv6 and IPv4 forwarding, and the SRv6 \textit{endpoint} behaviors, reported in Table \ref{tab:pe2_vpp}. The results of all the SRv6 behaviors are combined in Figure~\ref{fig:vpp_linux}. 

To discuss the results, we need to consider the line packet rate of the different behaviors in the configuration used in the experiments. Let us start from the \textit{policy headend} behaviors. For \textit{H.Enc.V6}, \textit{H.Enc.V4} and \textit{H.Enc.L2} we used inner packet of 64 bytes. The encapsulation adds  an outer IPv6 packet header of 40 bytes with no SRH header, because we configured the VPP node to add a single SRv6 segment (in this case the address of the single segment is simply carried in the IPv6 destination address). The resulting line packet rate for the encapsulated packet is $\approx$8803 kpps (see equation~\ref{eq:linerate}). Note that for \textit{H.Enc.L2} the inner Ethernet frame is 64 bytes, i.e. the inner IP packet is 50 bytes. For \textit{H.Insert} the incoming IPv6 packet is 64 bytes and the VPP node adds a segment list of two segments corresponding to an SRH header of 40 bytes. The resulting line packet rate is $\approx$8803 kpps also in this case.

For all the above mentioned SRv6 \textit{policy headend} behaviors, VPP does not reach the line packet rate, so we can appreciate the differences in the behavior performance. \textit{H.Insert} shows a lower performance with respect to the other behaviors (its PDR is $\approx$7387 kpps). In VPP, \textit{H.Insert} requires two memory copy operations: the first one to move the IPv6 header to create the space required for the SRH insertion, and the second one to copy the actual SRH into the created space. Instead, the other behaviors do not require the first memory copy operation as the SRv6 encapsulation is copied directly in the memory preceding the packet. Indeed, \textit{H.Encaps.V6} and \textit{H.Encaps.V4} exhibit respectively $\approx$7709 kpps and $\approx$8471 kpps, and \textit{H.Encaps.L2} is able to forward the traffic at $\approx$8052 kpps. As expected \textit{H.Encaps.V4} performs better than \textit{H.Encaps.V6}. Moreover, it is possible to appreciate a very low variability in Table \ref{tab:pe1_vpp}.

As for the SRv6 \textit{endpoint} behaviors, the line packet rate is $\approx$6868 kpps, considering an inner packet of 64 bytes, an outer IPv6 packet header of 40 bytes and a 40 byte SRH header with two segments (see equation~\ref{eq:linerate}). In our experiments, VPP is able reach the line packet rate for all the SRv6 \textit{endpoint} behaviors. Therefore, using our test methodology, we cannot evaluate the PDR of these behaviors for VPP. This is evident in Figure~\ref{fig:vpp_linux}, which shows the same PDR of $\approx$6868 kpps (corresponding to the line packet rate) for the 8 rightmost behaviors. Using 40GbE NICs instead of 10GbE ones would scale up by a factor 4 the line packet rate and should allow to hit the performance limit of VPP and evaluate the PDR for the \textit{endpoint} behaviors.

\section{Related Works}
\label{sec:related}

The software forwarding performance on commodity CPUs require careful measurement and analysis as such CPUs were not designed specifically for packet forwarding. In order to address these needs, several frameworks have been developed. However, none of the works found in literature have fully addressed the performance of SRv6 data-plane implementation either in the Linux kernel or in other software router implementations (e.g., VPP). Our previous work \cite{ahmedperformance} has started considering this topic focusing on the implementation in the Linux kernel.

DPDK~\cite{dpdk} is the state of the art technology for accelerating the virtual forwarding elements. It bypasses the kernel processing, balances the incoming packets over all the CPU cores and processes them in batches to make a better use of the CPU cache. In~\cite{begin2018accurate}, the authors presented an analytical queuing model to evaluate the performance of a DPDK-based vSwitch. The authors studied several characteristics of the DPDK framework such as average queue size, average sojourn time in the system and loss rate under different arrival loads.

In \cite{pitaev2017multi}, the performance of several virtual switch implementations including Open vSwitch (OVS)~\cite{ovs}, SR-IOV and VPP are investigated. The work focuses on the NFV use-cases where multiple VNFs run in x86 servers. The work shows the system throughput in a multi-VNF environment. However, this work considers only IPv4 traffic and does not address SRv6 related performance. In \cite{pitaev2018characterizing}, the previous work has been extended by replacing OVS with OVS-DPDK~\cite{ovs-dpdk}, which promises to significantly increase the I/O performance for virtualized network functions. They use DPDK-enabled VNFs and show how OVS-DPDK throughput compares to SR-IOV and VPP as the number of VNFs is increased under multiple feature configurations. However, the work still considers only plain IPv4 forwarding. 

In \cite{barach2018batched}, the authors explain the main architectural principles and components of VPP including: vector processing, kernel bypass, packets batch processing, multi-loop, branch-prediction and direct cache access. To validate the high speed forwarding capabilities of VPP, the authors reports some performance measurements such as packet forwarding rate for different vector sizes (i.e, number of packets processed as a single batch), the impact of multi-loop programming practice on the per-packet processing cost as well the variation of the packet processing rate as a function of the input workload process. However, this work analyses VPP forwarding performance only for plain IPv4 forwarding and does not consider other types of forwarding such as IPv6 and SRv6.

Open Platform for NFV Project (OPNFV)~\cite{opnfv} is a Linux foundation project which aims at providing a carrier-grade, integrated platform to introduce quickly new products and network services in the industry. The NFVbench~\cite{nfvbench} toolkit, developed under the  OPNFV umbrella, allows developers, system integrators, testers and customers to measure and assess the L2/L3 forwarding performance of an NFV-infrastructure solution stack using a black-box approach. The toolkit is agnostic of the installer, hardware, controller or the network stack used. VSPERF~\cite{vsperf} is another project within the OPNFV specialized for benchmarking virtual switch performance. VSPERF reported results for both VPP and OVS, which are based on daily executed series of test-cases~\cite{vsperf-results}. 

The FD.io project has released a technical paper~\cite{intel-cisco-report} for analysing the performance of several dataplane implementations such as DPDK, VPP, OVS-DPDK. The work reports a comparison between DPDK L2 forwarding, OVS-DPDK L2 Cross-Connect, VPP L2 Cross-Connect and VPP IPv4 forwarding in terms of throughput measured in pps. The FD.io Continuous System Integration and Testing (CSIT) project released a report characterizing VPP performance~\cite{csit-report}. The report describes a methodology to test VPP forwarding performance for several test cases including: L2 forwarding, L3 IPv4 forwarding, L3 IPv6 forwarding as well as some SRv6 behaviors. Regarding the latter, the report shows the performance of SRv6 \textit{H.Encaps}, \textit{End.AD}, \textit{End.AM} and \textit{End.As} behaviors. However, the report does not cover the performance of the rest of the SRv6 \textit{policy headend} and \textit{endpoint} behaviors. 
 
The performance of some SRv6 behaviors is reported in \cite{lebrun2017implementing}, \cite{lebrun2017reaping}. The work has mainly focused on some SRv6 \textit{policy headend} behaviors such as \textit{H.Insert} and \textit{H.Encaps}. The reported results show the overhead introduced by applying the SRv6 encapsulation to IPv6 traffic. However, the performance reported in this work can be considered out-dated as it considered the SRv6 implementations in Linux kernel 4.12 release. Moreover, the work does not report the performance of any SRv6 \textit{endpoint} behavior as they were not supported by the Linux kernel at that time. 

The work in \cite{vladislavic2019throughput} presents a performance evaluation methodology for Linux kernel packet forwarding. The methodology divides the kernel forwarding operations into execution area (EA) which can be pinned to different CPU cores (or the same core in case of single CPU measures) and measured independently. The EA are: i) sending; ii) forwarding; iii) receiving. The measured results considers only the OVS kernel switching in case of single UDP flow.

In \cite{mayer2019efficient}, the authors extends the SRv6 implementations in the Linux kernel to support the SRv6 dynamic proxy (\textit{End.AD}) behavior described in \cite{ietf-spring-sr-service-programming}. The authors name their proposal SRNK (SR-Proxy Native Kernel). The idea is to integrate the SRv6 dynamic proxy implementation described in \cite{8004208} directly in the Linux kernel instead of relying on an external kernel module. The work compares the performance of the SRv6 \textit{End.AD} behavior in SRNK implementation and SREXT \cite{srext-srv6-net-prog}.  

\cite{teamsegment} presents a solution where low-level network functions such as SRv6 encapsulation are offloaded to the Intel FPGA programmable cards. In particular, the authors partially offload the SRv6 processing from VPP software router to the NICs of the servers increasing data-path performance and at the same time saving resources. These precious CPU cycles are made available for the VNFs or for other workloads in execution on the servers. The work compares the performance of the SRv6 \textit{End.AD} behavior executed by a pure VPP implementation and by an accelerated solution. Tests results show in the worst scenario that the FPGA cards bring a CPU saving of 67.5\%. Moreover, the maximum throughput achievable by a pure VPP solution with 12 cores is achieved by the accelerated solution using only 6 cores.

\cite{leeperformance} studies SRv6 as alternative user plane protocol to GTP-U \cite{gtpu}. Firstly, authors proposes an implementation of the GTP-U encap/decap functions and of the SRv6 stateless translation behaviors defined in \cite{id-srv6-mobile-uplane}. These behaviors guarantee the coexistence of the two protocols which is crucial for a gradual roll-out. Authors used programmable data center switches to implement these dataplane functionality. Since it is hard to get telemetry from commercial traffic generator when a translation takes place, authors injected timestamp with a resolution of nanoseconds to measure the latency of SRv6 behaviors. Finally, they measured throughput and packet loss under light and heavy traffic conditions on a local environment. Results show no huge performance drop due to the SRv6 translation. Moreover, the latency of the SRv6 behaviors is similar to the GTP-U encap/decap functions.

In \cite{ahmedperformance}, the authors reports the performance of some SRv6 \textit{policy headend} and \textit{endpoint} behaviors. The work focuses on the Linux kernel and shows the performance of the SRv6 behaviors in comparison to plain IPv6 forwarding (IPv4 related behaviors have not been considered). The work analyses some performance issues of the SRv6 implementations in the Linux kernel related cross-connect behaviors. However, it does not provide any solution to fix these performance issues. Moreover, \cite{ahmedperformance} does not considered the SRv6 implementations in other software router implementations such as VPP. The work described in this paper extends and completes the work started in \cite{ahmedperformance} in several directions. Firstly, VPP has been integrated into the SRPerf platform and its performance evaluation is reported. \cite{ahmedperformance} reports the performance of the Linux kernel 4.18 while this work considers Linux kernel 5.2 \textit{net-next} and also IPv4 related SRv6 behaviors.

With respect of \cite{ahmedperformance}, this work improved the PDR finding procedure. Firstly, the previous procedure described in \cite{ahmedperformance} required a per-forwarder tuning in order to correctly set a minimum lower bound for the rate. The new PDR finder does not require any tuning of the initial rate lower bound. Indeed, the previous algorithm results to be less efficient - this can be noticed particularly with forwarders that can match the line rate of the Traffic Generator. Let us discuss briefly the complexity of the algorithms with an analysis of the worst case scenario. We assume an initial traffic rate of $1\%$ of the line packet rate and a target accuracy of $1\%$ of the line packet rate. The old PDR finder performs first a search with an exponential increase: it starts from $1\%$ and doubles the rate value at each iteration. It requires at most $ log_2(100) \leq log_2(128) = 7 $ steps before terminating with a search window about $[64,100]$. At this point the logarithmic/binary search phase starts, which takes at most 6 steps to reduce up to $1\%$ of the line packet the size of the search window (the binary search requires $ log_2(36) < log_2(64) = 6 $ steps). Instead the new algorithm performs directly a binary search on the interval $[1,100]$, this requires in the worst case 7 steps to get into desired state ($ log_2(100) \leq log_2(128) = 7 $).

Finally, this work addresses the performance problems regarding \textit{End.DX6} and \textit{End.X} identified in \cite{ahmedperformance}. Moreover, it introduces and evaluates the \textit{End.DT4} behavior which is currently missing in the Linux kernel.

\section{Conclusions}
\label{sec:conclusions}

In this paper, we have described the design and implementation of SRPerf, a performance evaluation framework for SRv6 implementations. SRPerf has been designed to be extensible: it can support different forwarding engines including software and hardware forwarding, but can also be extended to support different traffic generators. For example in this work we have shown the integration of the VPP software forwarding engine. Moreover, we have presented our evaluation methodology for the forwarding performance based on the estimation the PDR (Partial Drop Rate) metric.

We have used SRPerf to evaluate the performance of the most used SRv6 behaviors in the Linux kernel and in VPP. The results concerning the Linux kernel implementation show reasonable performance and no particular issues have been observed once we fixed some problems that were already identified in \cite{ahmedperformance}. As regards VPP, it is possible to obtain higher forwarding rates and for the \textit{endpoint} behaviors we actually reach the line packet rate for a 10GbE interface cards. The difference in the results between Linux and VPP was expected, since VPP leverages DPDK to accelerate the packet forwarding and a comparison between VPP and Linux is not fair at the moment. 
The SRPerf tool is valuable for different purposes, like support for the development and validation of new behaviors or testing and detection of issues in the existing implementations. It fills a gap in the space of reference evaluation platforms to test network stacks. In this respect, we have shown how we have used SRPerf to identify and fix two issues found in the SRv6 implementations of two cross-connect behaviors. Moreover, the \textit{End.DT4} behavior was missing and we have added it to the Linux kernel. We plan to contribute back these behaviors to the Linux kernel.

Finally, directions for future work include evaluating the performance of the SRv6 behaviors not addressed in this paper such as mobile user-plane behaviors and SRv6 proxy behaviors used in service function chaining (SFC) use-cases (see Table \ref{tab:support}). We are also working to improve the Linux kernel implementation of SRv6, considering eXpress Data Path \cite{xdp} (XDP). This framework provides packet processing at the lowest point in the Linux kernel stack and thus allows avoiding most of the overhead introduced by the Linux kernel. We plan to include XDP based SRv6 processing in our framework and perform an experimental analysis.

\section*{Acknowledgment}
    This work has received funding from the Cisco University Research Program Fund.
    




%

\bibliographystyle{IEEEtran}
\bibliography{main}

\end{document}